\documentclass[a4paper]{spie}  

\usepackage[]{graphicx}

\renewcommand {\d}  {{\rm d}}
\renewcommand {\i}  {{\rm i}}

\newcommand {\ee}  {{\rm e}}
\newcommand {\ch}  {{\rm ch}}

\newcommand {\E}  {{\varepsilon}}
\newcommand {\om}  {{\omega}}
\newcommand {\Om}  {{\Omega}}

\newcommand {\cD}  {{\cal D}}

\newcommand {\cF}  {{\cal F}}

\newcommand {\cN}  {{\cal N}}

\newcommand {\teta}  {{\tilde{\eta}}}

\newcommand {\Nu}  {{N_{\rm u}}}
\newcommand {\Nd}  {{N_{\rm d}}}

\newcommand {\lamu}  {{\lambda_{\rm u}}}
\newcommand {\kp}  {{\kappa}}
\newcommand {\kpd}  {{\kp_{\rm d}}}
\newcommand {\kpa}  {{\kp_{\rm a}}}

\newcommand {\Lbar}  {{\bar{L}}}
\newcommand {\barkpd}  {{\bar{\kp}_{\rm d}}}
\newcommand {\Ld} {{L_{\rm d}}}
\newcommand {\La} {{L_{\rm a}}}
\newcommand {\Lu} {{L_{\rm u}}}

\newcommand {\coul}  {\Lambda}
\newcommand {\atf}   {{a_{\rm TF}}}
\newcommand {\dUmax} {U^{\prime}_{\max}}

\title{Number of photons and brilliance  of the radiation from a
 crystalline undulator}

\author{
A.~V.~Korol\supit{a}, A.~V.~Solov'yov\supit{b} 
and
W.~Greiner\supit{b}
\skiplinehalf
\supit{a} Department of Physics, Russian Maritime Technical University,
          Leninskii prospect 101, St.~Petersburg  198262, Russia; \\
\supit{b} Frankfurt Institute for Advanced Studies,
          Johann Wolfgang Goethe-Universit\"at,
          60054 Frankfurt am Main, Germany
}  
\authorinfo{
Further author information: A.V.S. is
on leave from Ioffe Physical-Technical Institute,
Russian Academy of Sciences, Polytechnicheskaya 26,
St.~Petersburg 194021, Russia.}

\begin{document}
  \maketitle 

\begin{abstract}
The scheme for accurate quantitative treatment of the
radiation from a crystalline undulator in presence of the 
dechanneling and the photon attenuation is presented.
The number of emitted photons and the brilliance of 
electromagnetic radiation generated 
by ultra-relativistic positrons channeling in a crystalline undulator 
are calculated for various crystals, positron energies and 
different bending parameters.
It is demonstrated that with the use of high-energy 
positron beams available at present in modern colliders
it is possible to generate the crystalline undulator radiation 
with energies from hundreds of keV up to tens of MeV region.
The brilliance of the undulator radiation within this energy range
is comparable to that of conventional light sources of the third
generation but for much lower photon energies.

\end{abstract}

\keywords{crystalline undulator, dechanneling, photon attenuation, brilliance}

\section{INTRODUCTION} \label{Introduction}
  
In this paper new results from the theory of
electromagnetic radiation emitted by a bunch of 
ultra-relativistic positrons channeling through 
a periodically deformed crystal (a crystalline
undulator) are reported.
We formulate the approximation for effective analytical and
numerical analysis of the characteristics of the undulator radiation
with account for the influence of two main parasitic effects, the
positron dechanneling and the photon attenuation.
The developed formalism is applied to calculate the number of the emitted
photons and the brilliance of the radiation formed in crystalline
undulators.

In a crystalline undulator there appears, in addition to a well-known 
channeling radiation, the radiation of an undulator type 
which is due to the periodic motion of channeling particles 
which follow the bending of the crystallographic planes. 
The parameters of the undulator radiation can be easily varied by 
changing the energy of beam particles and the parameters of crystal bending.
The feasibility of this scheme was explicitly demonstrated for
in 
 Refs.~\citenum{KSG1998,KSG1999}.
In these papers as well as in the subsequent publications
\cite{KSG00Loss,KKSG00Tot,KSG01a,GKS01b,KKSG01c,KKSG02}
the idea of this new type of
radiation, the essential conditions and limitations which must be
fulfilled to make possible the observation of the effect were 
formulated in an adequate form for the first time. 
A number of corresponding numerical results were presented to 
illustrate the developed theory.
The importance of the ideas suggested and discussed in the cited papers
has also been realized by other authors resulting in a significant
increase of the number of publications in the field during last
years
\cite{Arm98,MU2000,Arm01a,Arm02,Arm03,B03,B04,Arm04,Arm05}
but, unfortunately, often
without proper citation 
\cite{Arm01a,Arm02,Arm03,B03,B04,Arm04,Arm05}.
A detailed review of the results obtained in this newly 
arisen field as well as a historical survey of the development of all
principal ideas and related phenomena can be found in 
Ref.~\citenum{KSG2004_review}.
\begin{figure}
\begin{center}
\includegraphics[width=13cm,height=5.5cm,angle=0]{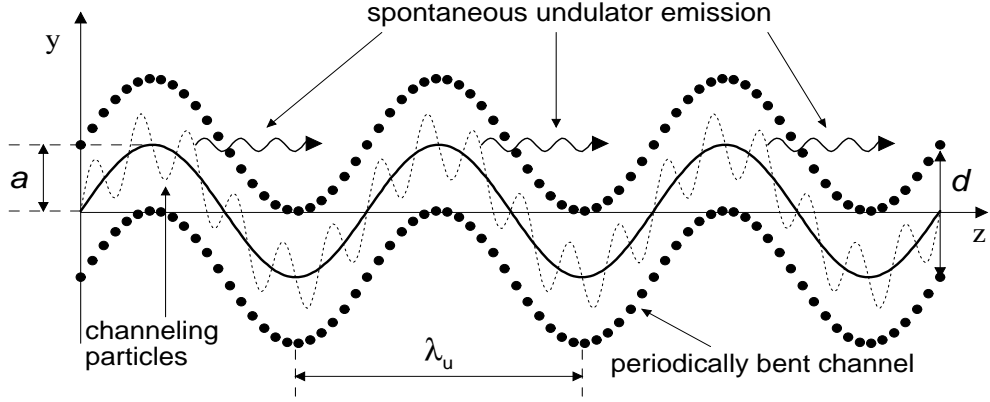}
\end{center}
\caption{Schematic representation of a crystalline undulator.
Two black circles denote the nuclei belonging to 
two neighbouring crystallographic planes 
(separated by an interplanar distance $d$)
which are periodically bent.
The centerline of this channel (solid line) is described by
a harmonic function $y(z) = a\sin(2\pi z/\lamu)$.
Its period $\lamu$ and amplitude $a$ satisfy the condition
$\lamu \gg a$.
The dashed curve represents the trajectory of a projectile
trapped in the channel.  
}
\label{figure1_jpg}
\end{figure}

The mechanism of the photon emission by means of a crystalline
undulator is illustrated in Fig.~\ref{figure1_jpg}.  
The $(yz)$-plane in the figure is a cross section of 
an initially linear crystal,
and the $z$-axis represents the cross section of a midplane of two
neighbouring non-deformed crystallographic planes (not drawn in 
the figure) spaced by the interplanar distance $d$.

Under certain conditions the ultra-relativistic positrons
will channel in the periodically bent channel.
The trajectory of a particle contains two elements.
Firstly, there are channeling oscillations due to the action of
the interplanar potential.
Their typical frequency $\Om_{\ch}$ depends on 
the positron energy $\E$ and the parameters of the 
interplanar potential.
Secondly, there are oscillations related to the periodicity of the 
distorted midplane, - the undulator oscillations, whose
frequency is $\om_0=2\pi c/\lamu$.

The spontaneous emission of photons is associated  with both of 
these oscillations. 
The typical frequency of the channeling radiation
is $\om_{\ch}\approx 2\gamma^2\Om_{\ch}$ where $\gamma=\E/mc^2$ 
is the relativistic Lorenz factor.
The undulator oscillations give rise to the photons with  
frequency $\om_{\rm u} \approx 4\gamma^2\om_0/(2+p^2)$ where $p$ is 
the undulator parameter,  $p=2\pi\gamma (a/\lamu)$ 
\cite{KSG1998,KSG1999}.
If strong inequality $\om_0 \ll \Om_{\ch}$ is met than the 
frequencies of the channeling radiation and the undulator 
radiation are also well separated, $\om \ll \om_{\ch}$.
In this case the characteristics of the undulator radiation 
are practically 
independent on the channeling oscillations but depend on the shape 
of the periodically bent midplane
\cite{KSG00Loss,KKSG00Tot}.


There are essential features which distinguish a crystalline
undulator from a conventional  one based on the action of the
periodic magnetic (or electric) field on the projectile.
In the latter the beam of particles and the photon flux move in vacuum 
whereas in the proposed scheme they propagate through a crystalline media.  
Therefore, to prove that the crystalline undulator 
is feasible, 
it is necessary to analyze the influence of 
the interaction of both beams with the crystal constituents.
On the  basis of such analysis one can formulate the 
conditions which must be met and define the ranges of parameters 
(which include $\E$, $a$, $\lamu$ and also
the crystal length $\Lu$ and the photon energy $\hbar\om$) 
within which all the criteria are fulfilled. 
In full this analysis was carried out very recently
and the feasibility of the crystalline undulator 
was demonstrated in an adequate form for the first time in 
Refs.~\citenum{KSG1998,KSG1999}
and in 
Refs.~\citenum{KSG00Loss,KKSG00Tot,KSG01a,GKS01b,KKSG01c,KKSG02}.

For further referencing let us briefly mention the conditions 
which must be met in a crystalline undulator.

A stable planar channeling of an ultra-relativistic positron
in a periodically bent crystal occurs if the maximum centrifugal force,
$F_{\rm cf}$, 
is less than the maximal force due to the interplanar field,
$F_{\rm int}$.
Notating the ratio $F_{\rm cf}/F_{\rm int}$ as $C$ one
formulates this condition as follows
\cite{KSG1998,KSG1999,BarDubGr80}:
\begin{eqnarray}
C =(2\pi)^2{\E \over U_{\max}^{\prime}} {a\over \lambda_{\rm u}^2} \ll 1\, .
\label{Condition.1}
\end{eqnarray}

There are two essentially different regimes of the radiation formation 
in a periodically bent crystals.
They are defined by the magnitude of the ratio $a/d$.
In the case of low amplitudes, $a/d\ll 1$, the characteristic
frequencies of the channeling radiation  and 
the undulator radiation become
compatible $\om_{\rm u}\sim \om_{\ch}$. 
This results in the loss of the monochromaticity of the radiation,
since the channeling radiation is essentially non-monochromatic
due to noticeable deviations of the interplanar potential from a harmonic
form.
Additionally, 
in this case the intensity of undulator radiation is small compared
with that of the channeling radiation
\cite{KSG00Loss,KKSG00Tot}.

On the contrary, in the limit $a\gg d$ not only 
the characteristic frequencies are well separated, 
$\om_{\rm u}/\om_{\ch}\approx C\, d/a \ll 1 $, 
but also the undulator radiation intensity is higher than 
the intensity of the channeling radiation 
\cite{KSG1998,KSG1999,KKSG00Tot}.
As a result, if one is only interested in the spectral distribution of the
undulator radiation, one may disregard the channeling oscillations and 
assume that the projectile moves along the centerline of the 
bent channel. 
Therefore, the criterion which is imposed on the relative magnitudes
of $d$, $a$ and $\lamu$ is as follows 
\begin{eqnarray}
d \ll a \ll \lamu\, .
\label{Condition.2}
\end{eqnarray}
The second inequality ensures that the crystal is deformed
elastically, and its structure and symmetry are not affected
by the deformation.

The term 'undulator' implies that the number of undulator periods, $\Nu$,
is large. 
Only in this limit does the radiation formed during the passage of a 
bunch of relativistic particles through a periodic system 
bear the features of undulator radiation 
(narrow, well-separated peaks in spectral-angular distribution)
rather than those of synchrotron radiation.
Hence, the following strong inequality, which 
entangles the period $\lamu$ and the length of a crystal
$\Lu$ must be met in the crystalline undulator 
\cite{KSG1998,KSG1999}:
\begin{eqnarray}
\Nu = {\Lu \over \lamu} \gg 1.
\label{Condition.3}
\end{eqnarray}

The coherence of the radiation, emitted in the crystalline undulator, 
takes place if the energy of the channeling particle does not 
change noticeably during with the penetration distance.
For ultra-relativistic projectiles the main source of energy losses
are the radiative losses.
Therefore, it is important to establish the range of energies 
for which the parameters of undulator radiation formed
in a perfect periodic crystalline structure are stable.
In Ref.~\citenum{KSG00Loss} a comprehensive quantitative
analysis of the radiative loss of energy, $\Delta \E$, due to the 
channeling and the undulator radiation was carried out. 
It was established that the relative radiative losses $\Delta \E/\E$ 
become large if the initial energy of the positron bunch is $\E > 10$ Gev. 
For lower energies of positrons
\begin{eqnarray}
\E \, <  10\, \mbox{GeV},
\label{Condition.4}
\end{eqnarray}
the radiative losses are small, $\Delta \E < 0.01 \E$.

As was pointed out Refs.~\citenum{KSG1998,KSG1999,KSG01a,GKS01b,KKSG01c}
two phenomena, the dechanneling effect and the photon attenuation, 
lead to severe limitation on the length
of a crystalline undulator.

If the  dechanneling effect is neglected, one may unrestrictedly
increase the intensity of the undulator radiation by considering larger
$\Nu$-values. 
In reality, random scattering of the channeling particle by the 
electrons and nuclei of the crystal leads to a gradual increase of 
the particle energy associated with the transverse oscillations 
in the channel. 
As a result, the transverse energy at some distance from the
entrance point exceeds the depth of the interplanar potential well,
and the particle leaves the channel. 
The mean penetration distance covered by a channeling particle is called 
the dechanneling length.
For given crystal and projectile the dechanneling
length $\Ld=\Ld(\E,C)$ depends on the energy $\E$  
and on the parameter $C$ (see (\ref{Condition.1})).
To calculate the dechanneling length one can either apply the diffusion
theory to describe the multiple scattering \cite{Kudo1981,EllisonPicraux1981}
or carry out a computer simulation of the scattering process of 
the projectile from the crystal constituents 
\cite{KSG01a,BiryukovChesnokovKotovBook}.
Alternatively, to estimate $\Ld(\E,C)$ one can 
use the approximate formulae \cite{KSG1999,BiryukovChesnokovKotovBook}.
For an ultra-relativistic positron the dechanneling length in straight
channels (i.e. $C=0$) in various crystals lies within the 
interval 
$\Ld(\E,0)\, \mbox{(cm)} \approx (0.05\dots0.08) \, \E \, \mbox{(GeV)}$
\cite{KSG2004_review}, i.e. does not exceed several
millimeters at  GeV energies of a positron.
For a periodically bent channel the dechanneling length decreases as
$C$ grows following, approximately, 
the law $\Ld(\E,C)\approx (1-C)^2\Ld(\E,0)$
\cite{KSG01a,BiryukovChesnokovKotovBook}.

The propagation of photons emitted in a crystalline undulator
is strongly influenced by the atomic and the nuclear photoeffects, 
the coherent and incoherent scattering on electrons and nuclei, the
electron-positron pair production.
All these processes lead to the decrease in the intensity of the photon
flux as it propagates through the crystal.
A quantitative parameter, which accounts for all these effects and
defines the scale within which the intensity of a photon flux 
decreases by a factor of $e$,
is called the attenuation length,  $\La = \La(\om)$. 
It is related to the mass attenuation coefficient $\mu(\omega)$ as
$\La(\om)=1/\mu(\om)$ \cite{Hubbel,ParticleDataGroup2004}.
The coefficients $\mu(\omega)$ are tabulated for all elements and 
for a wide range of photon frequencies \cite{Hubbel}.
The magnitude of $\La(\om)$
depends on $\om$ and on the type of the constituent atoms.
For low-$Z$ crystals (e.g., diamond) the magnitude of $\La(\om)$
exceeds that for a tungsten crystal taken for the same $\om$
by a factor of $10^1\dots 10^3$.
In the case of a diamond crystal the value of $\La(\om)$ varies from
$10^{-2}$ cm at $\hbar \om \approx 5$ keV up to several cm for
$\hbar \om > 10^5$ eV.

The simplest way to account for the dechanneling  
and the attenuation is to consider the case when the crystal
length satisfies the condition 
$\Lu < \min\left[\Ld(\E,C),L_a(\om)\right]$, and to assume
that within the chosen $\Lu$ scale neither the number of channeled
particles nor the flux of emitted photons do decrease.
Such approach was utilized in most of the papers devoted to 
the crystalline undulator problem.
More consistent treatment of the dechanneling process and of its influence
on the parameters of the undulator radiation was carried out in 
Refs.~\citenum{KSG01a,GKS01b},
where a simple analytic expression for the spectral-angular distribution
was derived which contains, as a parameter, the dechanneling length.
      
In the present work we make another step in developing the theory
of the crystalline undulator.
The following problems are solved and discussed below in the paper.
\begin{itemize}
\item[(a)]
We propose the scheme for accurate quantitative treatment of the
radiation from a crystalline undulator in presence of the 
dechanneling and the photon attenuation 
(Sect.~\ref{CrystallineUndulator}). 
As a result, we evaluate an analytic expression for the
spectral-angular distribution and the number of emitted photons
which contains, as parameters, three quantities 
$\Lu$, $\Ld(\E,C)$, and $L_a(\om)$.

\item[(b)]
We demonstrate that for given type of the crystal and crystallographic 
plane, and for given values of $\E$, $a$, $\lamu$ and $\om$ there
exists an optimal length of the crystal which ensures the largest 
number of the emitted photons.

\item[(c)]
Using (a) and (b) we carry out the calculation of 
 the number of emitted photons and the brilliance of the 
crystalline undulator radiation  (Sect.~\ref{NumericalResults}).
The calculations, which account for the conditions 
(\ref{Condition.1})-(\ref{Condition.4}),
 are performed for several crystals and by
using the parameters of positron bunches used in modern
colliders \cite{ParticleDataGroup2004}.

\end{itemize}

Prior to the discussion of the radiation formed in a crystalline
undulator, in  Sect.~\ref{IdealUndulator} we briefly 
summarize the results from the general theory 
of undulator radiation (see, e.g. 
Refs.~\citenum{Alferov,Kim1989,RullhusenArtruDhez}). 
We use the term `ideal undulator' to indicate that the propagation of 
positrons and photons occurs in vacuum. 

\section{Characteristics of radiation formed in an ideal undulator}
\label{IdealUndulator}

The spectral-angular distribution of the energy $E$ emitted
by an ultra-relativistic projectile in a planar undulator
can be written in the following form
\begin{eqnarray}
{\d^3 E \over \hbar \d\omega\,\d\Omega }
=
S(\om,\theta,\varphi)\,
 D_{N_{\rm u}}(\teta)\, .
\label{dspectral.1}
\end{eqnarray}
Here $\theta\ll 1$ and $\phi$ are the emission angles 
with respect to the undulator axis,
$\d\Omega = \theta \d \theta \d \varphi$ is the solid angle of the emission.
The function $S(\om,\theta,\varphi)$, which does not depend 
on the undulator length, is given by
\begin{eqnarray}
S(\om,\theta,\varphi)
=
{\alpha \over 4\pi^2}\,
{\omega^2 \over \gamma^2\omega_0^2}
\left\{
p^2\left|I_1\right|^2
+
\gamma^2\theta^2 \left|I_0\right|^2
-
2p\gamma\, \theta \cos \varphi\, {\rm Re}\left(I_0^{*} I_1\right) 
\right\}\, ,
\label{dspectral.2a}\\
 I_m
=
\int_0^{2\pi}
\d \psi\,
\cos^m\psi \,
\exp\left(
\i \left[\eta \psi 
+ {p^2\omega \over 8\gamma^2\omega_0}\,\sin(2\psi) 
- {p \omega \over \gamma\omega_0}\,\theta\cos\varphi\sin \psi
\right]
\right),
\qquad
m=0,1\,.
\label{dspectral.3}
\end{eqnarray}
Here $\alpha \approx 1/137$, 
$\om_0= 2\pi c/\lambda_{\rm u}$,
$p$ is the undulator parameter and the parameter 
$\eta$ is given by
 \begin{eqnarray}
\eta = {\omega \over 2 \gamma^2\omega_0}\,
\left(1 + \gamma^2\theta^2 + {p^2 \over 2}
\right).
\label{dspectral.4}
\end{eqnarray}
The factor  $D_{N_{\rm u}}(\teta)$ on the right-hand side of 
(\ref{dspectral.1}) is defined as follows
 \begin{eqnarray}
D_{\Nu}(\teta)
=
\left({\sin \Nu\pi \teta \over \sin \pi \teta}  \right)^2,
\label{IdealUndulator.2a}
\end{eqnarray}
where $\teta = \eta-n$ and  $n$ is a positive integer such that
$n-1/2< \eta \leq n+1/2$.

For $\Nu\gg 1$ the function $D_{\Nu}(\teta)$ has a sharp and powerful 
maximum  in the point $\teta=0$, where  $D_{N_{\rm u}}(0)=N_{\rm u}^2$. 
The width of the peak  $\Delta\teta_{\rm u}$ is equal to $1/\Nu$.
This behaviour of $D_{\Nu}(\teta)$ results in a peculiar form
of the spectral-angular distribution of undulator radiation
which clearly distinguishes it from other types of electromagnetic 
radiation formed by a charge moving in external fields.
Namely, for each value of the emission angle $\theta$
the spectral distribution
consists of a set of narrow and equally spaced peaks (harmonics).
The peak intensity is proportional to  $N_{\rm u}^2$.
This factor reflects the constructive interference of radiation emitted
from each of the undulator periods and is typical for any system which
contains $\Nu$ coherent emitters.

The values $\om_n$ of the harmonics frequencies follow from the
condition that parameter $\eta$ becomes
an integer (this corresponds to $\teta=0$).
In particular, in the case of the forward emission, $\theta=0$,
the harmonics frequencies are defined from the relation
\begin{eqnarray}
n = {1 \over 2 \gamma^2}{\omega_n \over \omega_0}\,
\left(1 + {p^2 \over 2}\right)\,.
\label{CentralLine.1}
\end{eqnarray}
For $\theta=0$ the integrals  (\ref{dspectral.3}) can be evaluated 
analytically and the spectral-angular distribution 
calculated for $\om=\om_n$ (for $n=1,3,5\dots$) acquires the form
\cite{Alferov,Kim1986}:
\begin{eqnarray}
\left.{\d^3 E \over\hbar\d\omega\,\d\Omega }
\right|_{\theta=0 \atop \om=\om_n}
=
\alpha \,N_{\rm u}^2\,\gamma^2\,
{n^2p^2  \over (1+p^2/2)^2}
\left[
J_{n-1 \over 2}(z)
-
J_{n+1 \over 2}(z)
\right]^2\, ,
\label{NumberPhotonsCollection.5a}
\end{eqnarray}
where $z=n p^2/(4+2p^2)$ and $J_{\nu}(z)$ is the Bessel function.

The finite width of the central peak of $D_{\Nu}(\teta)$
defines the emission cone $\Delta\Om_n$ and the bandwidth 
$\Delta\om_n/\om_n$ of the $n$th harmonic.
Using $\Delta\teta_{\rm u}=1/N_{\rm u}$  and accounting for 
(\ref{CentralLine.1}) one derives
\begin{eqnarray}
\Delta\Om_n 
=
{\pi \over \gamma^2}\,{ 1 + p^2/2 \over n N_{\rm u}}\,,
\qquad\qquad
{\Delta \om_n\over \om_n}
=
{1 \over nN_{\rm u}}\,.
\label{NumberPhotonsCollection.3}
\end{eqnarray}

Formulae (\ref{NumberPhotonsCollection.5a})-(\ref{NumberPhotonsCollection.3})
allow one to calculate the number of photons $\Delta N_{\om_n}$ 
of energy $\om=\Bigl[\om_n-\Delta\om_n/2,\om_n+\Delta\om_n/2\Bigr]$
emitted by a beam particle within the cone $\Delta\Om_n$:
\begin{eqnarray}
\Delta N_{\om_n}
=
\left.{\d^3 E \over\hbar\d\omega\,\d\Omega }\right|_{\theta=0 \atop \om=\om_n}
\!\!\Delta\Om_n \, {\Delta \om_n\over \om_n}
=
\pi\alpha\, N_{\rm u}\,
Q_n(p)\, {\Delta \om_n\over \om_n}
\label{NumberPhotonsCollection.1}
\end{eqnarray}
where 
$Q_n(p)=4z\left[J_{(n-1)/2}(z)-J_{(n+1)/2}(z)\right]^2$.
 
Let us introduce two other quantities which
characterize the radiation formed in an undulator and are closely
related to the number of the emitted photons, but also take into account
the properties of the beam of ultra-relativistic particles. 
These quantities are the {\it flux} and the {\it  brilliance} 
(see, e.g. Ref.~\citenum{RullhusenArtruDhez}).

The flux $F_n$ describes the number of 
photons per second of the $n$th
harmonic emitted in the cone $\Delta\Om_n$ and in a given bandwidth.
A quantitative definition of this quantity, measured in 
$\Bigl(\mbox{photons}/\mbox{s}/ 0.1\% \mbox{BW}\Bigr)$
(the abbreviation 'BW' stands for the bandwidth $\Delta \om_n/ \om_n$),
is given by the following formula
 \cite{Kim1986}:
\begin{eqnarray}
F_n
=
{\Delta N_{\om_n}\over 10^{3}(\Delta\om_n/\om_n)}\,
{I \over e}
=
10^{-3}\pi\alpha\, N_{\rm u}\,
Q_n(p)\, 
{I \over e}
=
1.431\times10^{14}\,N_{\rm u}\, Q_n(p)\,I\, \mbox{[A]}
\, ,
\label{B&B.4}
\end{eqnarray}
where $I$ is the electric current of the beam.
In the latter expression $I$ is measured in Amperes.

The general definition of brilliance 
of the photon source of a finite size 
is given in terms of the number of photons of energy $\hbar\om$ emitted
in the cone $\Delta\Om$ per unit time interval, unit source area,
unit solid angle and per bandwidth \cite{RullhusenArtruDhez}.  
To calculate this quantity is it necessary to know
the beam sizes $\sigma_x$, $\sigma_y$ and angular divergencies 
$\phi_{x}$, $\phi_{y}$ in two perpendicular directions, as well as
the divergency angle of the radiation and the 'size' of the photon
beam. 
The brilliance of undulator radiation can be  related to the flux $F_n$ 
as follows \cite{Kim1986}:
\begin{eqnarray}
B_n
=
{F_n \over (2\pi)^2\, \epsilon_x\epsilon_y}\, .
\label{B&B.7}
\end{eqnarray}
Here 
$\epsilon_{x,y}=\sqrt{\sigma_n^2+\sigma_{x,y}^2}\,\sqrt{\phi_n^2+\phi_{x,y}^2}$
are the total emittance of the 
photon source in the $x$ and $y$ directions, 
with  $\phi_n=\sqrt{\Delta\Om_n/2\pi}$ being the angular 
width of the $n$th harmonic and
$\sigma_n=\lambda_n/4\pi\phi_n$ is the 'apparent' source size
taken in the diffraction limit \cite{Kim1986NIM2}.

To obtain brilliance in the units
$\Bigl(\mbox{photons}/\mbox{s}/\mbox{mrad}^2/\mbox{mm}^2/0.1\%\mbox{BW}\Bigr)$ 
the quantities $\sigma_{x,y}$ and $\sigma_n$
must be measured in millimeters whereas
the angular variables  $\phi_{x,y}$ and $\phi_n$ -
 in milliradians.

\section{Characteristics of radiation formed in a crystalline undulator} 
\label{CrystallineUndulator}

In an ideal undulator the beam of particles and the emitted
photons propagate in vacuum.
In a crystalline undulator, due to the interactions with crystal
atoms, the particles can dechannel, and thus be lost for 
further motion through the undulator.
Additionally, the photons emitted inside the crystal can be absorbed
or scattered while making their way out from the crystal.
Therefore, it is necessary to account for the processes
of dechanneling and photon attenuation.
In what follows we carry out the qualitative analysis
of the influence of these two processes on
the characteristics of the radiation formed in a crystalline
undulator.

\subsection{Spectral-angular distribution 
in presence of the dechanneling and attenuation}
\label{DechannelingAndAttenuation}

Let the crystal length, the amplitude and period of bending
and the energy $\E$ satisfy 
the conditions (\ref{Condition.1})-(\ref{Condition.4}).

A positron, which enters the crystal at small
incident angle with respect to the curved crystallographic plane,
penetrates through the crystal following the bending of its channel.
However, due to random scattering by the electrons and nuclei of 
the crystal the energy of the transverse oscillations of the positron
in the channel increases, and finally the particle leaves the channel, 
becoming lost for the crystalline undulator.
Although the rigorous treatment of the dechanneling
process cannot be implemented by analytical means only, 
it is possible to develop a model approach based on the 
assumption that the probability $w(z)$ for a particle to penetrate at
a distance $z$ along the undulator axis ($z\in [0,\Lu]$)
can be described by the exponential decay law
\begin{eqnarray}
w(z) = \exp\left(-{z/ \Ld}\right)\, .
\label{DechannelingInUndulator.1}
\end{eqnarray}
In intermediate formulae when referring to the dechanneling length 
we omit its arguments $\E$ and $C$.

With the effect of dechanneling taken into account the
spectral-angular distribution of the radiated energy 
per one particle can be written as follows:
\begin{eqnarray}
{\d^3 E \over \hbar \d\om\,\d\Om }
=
{\d^3 E^{(-)} \over \hbar \d\om\,\d\Om }
+
{\d^3 E^{(+)} \over \hbar \d\om\,\d\Om }
\,.
\label{DechannelingInUndulator.6a}
\end{eqnarray}
The first term is the contribution to from all the processes
in which the particle dechannels somewhere inside the crystal.
To calculate this term one notices that
the quantity $\Ld^{-1}\d z\exp\left(-{z/ \Ld}\right)$
defines the probability of a particle to channel through the distance
$z$ and then dechannel within the 
interval $\d z$.
Such a particle emits the radiation which corresponds to the undulator 
of the length $z$ and the number of periods $z/\lamu$.
Therefore
\begin{eqnarray}
{\d^3 E^{(-)} \over \hbar \d\om\,\d\Om }
=
\int_0^L{\d z \over \Ld}\,\ee^{-{z/\Ld}}\,
{\d^3 E^{(att)}(z) \over \hbar \d\om\,\d\Om }\, ,
\label{DechannelingInUndulator.4}
\end{eqnarray}
where $\d^3 E^{(att)}(z)/\hbar \d\om\,\d\Om$ 
denotes the spectral-angular distribution from the undulator of the length
$z$.
The superscript `(att)' indicates that to calculate this 
quantity one has to account for the photon attenuation.

The second term on the right-hand side of (\ref{DechannelingInUndulator.6a})
is due the process when the projectile channels through the whole 
length $\Lu$.
Its probability is given by the factor
$\exp\left(-{\Lu/ \Ld}\right)$. 
Therefore one can write
\begin{eqnarray}
{\d^3 E^{(+)} \over \hbar \d\om\,\d\Om }
=
\ee^{-{\Lu/\Ld}}\,
{\d^3 E^{(att)}(\Lu) \over \hbar \d\om\,\d\Om }\, .
\label{DechannelingInUndulator.5}
\end{eqnarray}

If the photon attenuation is neglected, then
to calculate (\ref{DechannelingInUndulator.5})
one uses (\ref{dspectral.1}) instead of 
${\d^3 E^{(att)}(\Lu) / \hbar \d\om\,\d\Om }$.
The integral in 
(\ref{DechannelingInUndulator.4})
is also evaluated with the help of (\ref{dspectral.1}) 
where one  substitutes $\Nu$ with $z/\lamu$.
Such approach was applied Ref.~\citenum{KSG01a} 
with the only difference
that in the cited paper to calculate (\ref{DechannelingInUndulator.4})
and (\ref{DechannelingInUndulator.5}) we used 
the discrete probabilities instead of the continuous distribution
function (\ref{DechannelingInUndulator.1}). 
The use of the latter implies that the dechanneling effect is small
over the scale of one undulator period and, therefore,
$\Ld \gg \lamu$.

Now let us turn to the derivation of the quantity 
${\d^3 E^{(att)}(z) / \hbar \d\om\,\d\Om }$ which 
is the spectral-angular distribution of 
radiation formed in the undulator of the length $z\leq \Lu$
in presence of the attenuation.
In the intermediate formulae we assume that 
the ratio $N_z=z/\lamu$ is an integer number which corresponds to
the number of periods in this undulator.
In the final formula this limitation will be omitted.
Throughout the text the notations $\Lu$ and $\Nu$ are reserved
for the length of the crystal and the number of undulator periods
within $\Lu$.

As mentioned above, if one neglects the photon attenuation effect,
the distribution ${\d^3 E^{(att)}(z) / \hbar \d\om\,\d\Om }$ 
is described by (\ref{dspectral.1}) where one substitutes $\Nu$ 
with $N_z$.
The only quantity in  (\ref{dspectral.1})
which depends on the number of undulator periods is the 
factor $D_{N_z}(\teta)$ defined in (\ref{IdealUndulator.2a}).
This factor appears in the formula for spectral-angular distribution 
as a result of squaring the modulus of a coherent sum of the 
amplitudes of electromagnetic waves emitted from spatially different 
but similar parts of the undulator.
In more detail, $D_{N_z}(\teta)$ is given by
$D_{N_z}(\teta)=
\left|\sum_{j=1}^{N_z}\exp\Bigl(\i kR_0-2\i \pi\teta j\Bigr) \right|^2$.
The argument $(kR_0-2\pi\teta j)$ ($k=\om/c$ is the wavenumber) 
stands for the phase of  the electromagnetic wave 
emitted within the $j$th period of the undulator 
and detected at some distant point $R_0$ from the undulator.
It is assumed that the quantities $\Lu$, $z$ and $R_0$
satisfy the relations: $z \leq \Lu \ll R_0$.
\begin{figure}
\begin{center}
\includegraphics[width=10cm,height=5.5cm]{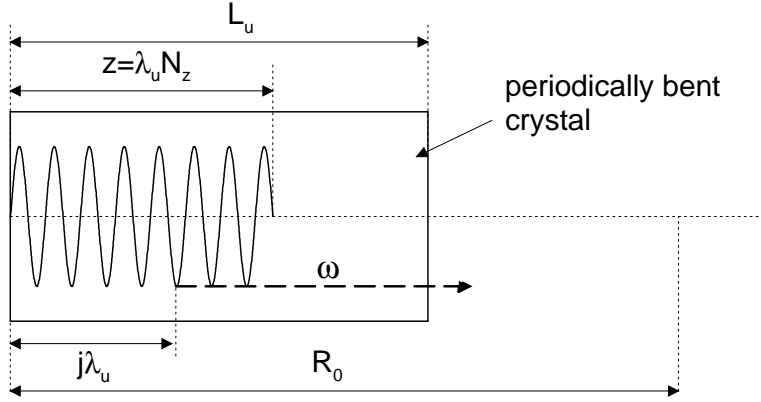}
\end{center}
\vspace{-0.5cm}
\caption{
Illustration of the photon attenuation in a crystalline undulator.
A photon (the long-dashed line), emitted within the $j$th period of
the undulator of the length $z$, can be absorbed (or scattered)
in the part of crystal of thickness $\Lu-j\lamu$ 
on its way to a distant detection point $R_0$ ($R_0\gg \Lu$).}
\label{ForDn.eps.fig}
\end{figure}

In a crystalline undulator a photon emitted within the $j$th period
in the direction of the point $R_0$ can be absorbed within the distance
$\Lu-j\lamu$ while propagating through the crystal,
see Fig.~\ref{ForDn.eps.fig}.
To account for this possibility one can assume that the wavenumber
becomes complex, $k\rightarrow \om/c + \i\mu/2$. 
The quantity $\mu=\mu(\om)$ defines the attenuation length 
$\La(\om)= \mu^{-1}(\om)$ within which the photon flux
is reduced by a factor of $e$.
For a complex $k$ the phase factor $\ee^{\i kR_0}$, which in an ideal 
undulator is the same for all periods $j=1\dots N_z$, 
is replaced with $\ee^{\i kR_0}\ee^{-\mu(\Lu-j\lamu)}$,
and a proper expression for $D_{\Nu}$ is 
 \begin{eqnarray}
D_{N_z}(\teta)
\rightarrow
D_{N_z}^{(att)}(\teta)
=
\left|
\ee^{\i kR_0}\sum_{j=1}^{N_z}\ee^{2\i \pi\teta j}
\ee^{-{\mu\over 2}(\Lu-j\lamu) }
\right|^2
=
\ee^{-\mu \Lu}
{
1+\ee^{\mu z}-2\ee^{\mu z/2}\cos(2\pi\teta  N_z)
\over 
1 + \ee^{\mu \lamu} - 2\ee^{\mu \lamu/2}\cos(2\pi\teta)}
\, ,
\label{AttenuationInUndulator.3}
\end{eqnarray}
The  spectral-angular distribution of radiation in presence
of the photon attenuation acquires the form
\begin{eqnarray}
{\d^3 E^{(att)}(z) \over \hbar \d\om\,\d\Om }
=
S(\om,\theta,\varphi)\, 
D_{N_z}^{(att)}(\teta)\, .
\label{AttenuationInUndulator.8}
\end{eqnarray}
In the limit $\mu\to 0$ (i.e., when there is no attenuation) 
the factor $D_{N_z}^{(att)}(\teta)$ becomes equal to $D_{N_z}(\teta)$ 
from (\ref{IdealUndulator.2a}), and the right-hand side of 
(\ref{AttenuationInUndulator.8}) reduces to that of 
Eq. (\ref{dspectral.1}).
 
To derive the explicit expression for the spectral-angular distribution 
of the radiated energy  from a crystalline undulator
one uses (\ref{AttenuationInUndulator.8}) in 
(\ref{DechannelingInUndulator.6a})--(\ref{DechannelingInUndulator.5}).
Let us note here, that although the expression 
(\ref{AttenuationInUndulator.3}) was obtained for the case when
the ratio $z/\lamu$ is an integer, its use in the integral
from (\ref{DechannelingInUndulator.4}) can be justified 
by the above mentioned conditions, 
that the undulator period $\lamu$ is small compared to the 
dechanneling length $\Ld$, the attenuation length, $\La$, and
the length of crystal, $\Lu$.
Hence, the relative error, which appears when one uses 
(\ref{AttenuationInUndulator.8}) in (\ref{DechannelingInUndulator.4}),
is small, being of the 
order of magnitude $\lamu/\min\{\Ld,\La,\Lu\} \ll 1$.
Carrying out the integration
one represents the total spectral-angular distribution 
(\ref{DechannelingInUndulator.6a}) of radiation formed in the
crystalline undulator
in the form similar to (\ref{dspectral.1})
\begin{eqnarray}
{\d^3 E \over \hbar \d\omega\,\d\Omega }
=
S(\om,\theta,\varphi)\,
 \cD_{\Nu}(\teta)\, ,
\label{DechannelingAndAttenuation.1}
\end{eqnarray}
where the function $S(\om,\theta,\varphi)$, defined by
(\ref{dspectral.2a}), does not depend on $\Lu$,  $\Ld$ and $\La$.
These parameters enter the factor $\cD_{\Nu}(\teta)$ which is
given by the expression:
\begin{eqnarray}
\cD_{\Nu}(\teta)
=
{4N_{\rm u}^2 \over \kp_{\rm a}^2 + 16N_{\rm u}^2\sin^2\pi\teta}
\Biggl[
{\kpa \over \kpa -\kpd}\,\ee^{-\kpd}
-
{2\kpd-\kpa\over \kpa -\kpd}
{\kp_{\rm a}^2+4\phi^2\over(2\kpd-\kpa)^2 +4\phi^2}\,
\ee^{-\kpa}
\nonumber\\
-
2
\left(
\cos\phi
+
2\kpd\,
{
2\phi\,\sin\phi-(2\kpd-\kpa)\cos\phi
\over
(2\kpd-\kpa)^2 +4\phi^2}
\right)
\ee^{-(2\kpd+\kpa)/2}
\Biggr]\, ,
\label{DechannelingAndAttenuation.2}
\end{eqnarray}
where the following notations are used:
\begin{eqnarray}
\kpd ={\Lu\over \Ld},
\qquad
\kpa ={\Lu\over \La},
\qquad
\phi=2\pi\teta \Nu\, .
\label{DechannelingAndAttenuation.3}
\end{eqnarray}

Despite a cumbersome form of the right-hand side of 
(\ref{DechannelingAndAttenuation.2}) its main features can be
easily understood. 
Firstly, we notice that if the dechanneling is neglected,
$\Ld\to\infty$ (or $\kpd\to 0$), 
the function $\cD_{\Nu}(\teta)$ reproduces $D_{\Nu}^{(att)}(\teta)$ 
from (\ref{AttenuationInUndulator.3}).
In another  limit $\kpd= \kpa=0$ 
(i.e., no attenuation and dechanneling)
eq. (\ref{DechannelingAndAttenuation.2}) reduces to the definition
of the factor $D_{\Nu}(\teta)$ which characterizes the ideal
undulator.
In the case when only the attenuation effect is neglected 
the limit of $\cD_{\Nu}(\teta)$ can also be easily evaluated. 
In either of these cases the main maximum of $\cD_{\Nu}(\teta)$ 
is located in the point $\teta=0$, i.e. when the parameter $\eta$ 
reduces to an integer, and,
therefore, the harmonics frequencies are still defined by 
(\ref{CentralLine.1}).
The maximum value $\cD_{\Nu}(0)$ can be presented as follows:
\begin{eqnarray}
\cD_{\Nu}(0)
=
4N_{\rm d}^2\,
\left[
{\ee^{-x\kpd}\over (1-x)(2-x)}
-{\ee^{-\kpd}  \over x(1-x)}
+
{2\ee^{-(2+x)\kpd/2}\over x(2-x)}
\right],
\label{DechannelingAndAttenuation.5}
\end{eqnarray}
where the quantity  $\Nd=\Ld/\lamu$ stands for the number of undulator 
periods within $\Ld$, and 
the ratio
\begin{eqnarray}
x={\kpa \over \kpd} = {\Ld \over \La}
\end{eqnarray}
does not depend on the crystal length $\Lu$.

The width of the central peak $\Delta\teta$, 
which in the case of an ideal undulator
equals to $1/\Nu$, is increased due to the photon attenuation 
and the dechanneling.
Formally, the additional widths are due to the factors
$1/(\kp_{\rm a}^2 + 16N_{\rm u}^2\sin^2\pi\teta)$ and 
$1/((2\kpd-\kpa)^2 +4\phi^2)$ 
which enter (\ref{DechannelingAndAttenuation.2}).
The widths associated with these factors are, respectively,
$\Delta\teta_1=\kpa/(2\Nu\pi)$ and 
$\Delta\teta_2=|2\kpd-\kpa|/(2\Nu\pi)$.
Thus, the total width of the peak is:
\begin{eqnarray}
\Delta\teta
=
\sqrt{N_{\rm u}^{-2}+(\Delta\teta_1)^2+(\Delta\teta_2)^2}
=
{1\over \Nu}
\sqrt{
1
+{(\kpa-\kpd)^2 +\kp_{\rm d}^2 \over 4\pi^2} 
}
\label{DechannelingAndAttenuation.6}
\end{eqnarray}
The additional widths 
lead to the enlargement of the solid angle $\Delta\Om_n$ 
of the emission cone in the forward direction.
In accordance with (\ref{DechannelingAndAttenuation.6}) 
one derives
\begin{eqnarray}
\Delta\Om_n 
=
{\pi \over \gamma^2}\,{ 1 + p^2/2 \over n \Nu}\,
\sqrt{
1
+\kp_{\rm d}^2\,{(x-1)^2 +1 \over 4\pi^2}\,.
}
\label{CUcharacteristics.1}
\end{eqnarray}

The formulae for the number of photons $\Delta \cN_{\om_n}$  
emitted in the cone $\Delta\Om_n$ as well
the corresponding flux of radiation $\cF_n$ one derives 
similarly to how it was done in Sect. \ref{IdealUndulator} for 
an ideal undulator.
The result is:
\begin{eqnarray}
\Delta \cN_{\om_n}
=
\pi\,\alpha \,N_{\rm eff}(x,\kpd)\,Q_n(p)\,{\Delta \om_n\over \om_n}\,
\label{CUcharacteristics.4a}\\
\cF_n
=
1.431\times10^{14}\,N_{\rm eff}(x,\kpd)\, Q_n(p)\,I\, \mbox{[A]}
\,.
\label{CUcharacteristics.4b}
\end{eqnarray}
The difference between these equations and formulae 
(\ref{NumberPhotonsCollection.1}) and (\ref{B&B.4}) is that 
the number of undulator periods $\Nu$, met in the latter,
is substituted with the effective number of periods, 
$N_{\rm eff}(x,\kpd)$, which is defined as follows:
\begin{eqnarray}
N_{\rm eff}(x,\kpd)
=
{\cD_{\Nu}(0) \over \Nu}\,
\sqrt{
1
+\kp_{\rm d}^2\,{(x-1)^2 +1 \over 4\pi^2} 
}
\equiv
\Nd\,f(x,\kpd)
\label{CUcharacteristics.5}\\
f(x,\kpd)
=
{4 \over \kpd}
\Biggl[
{\ee^{-x\kpd}\over (1-x)(2-x)}
-{\ee^{-\kpd}  \over x(1-x)}
+
{2\ee^{-(2+x)\kpd/2}\over x(2-x)}
\Biggr]
\sqrt{
1
+\kp_{\rm d}^2\,{(x-1)^2 +1 \over 4\pi^2} 
}\,.
\label{CUcharacteristics.6}
\end{eqnarray}
We use these equations in the subsequent section to
define the optimal length of a crystalline undulator.

\subsection{Optimal length of a crystalline undulator}
\label{OptimalLength}

In the case of an ideal undulator one can, in principle, 
increase infinitely the length of the undulator. 
This will result in the increase of the number of photons,
the photon flux, and the brilliance since they are proportional to 
the number of periods.
The limitations on the values of $\Lu$ and $\Nu$  are
mainly of a technological nature.

The situation is different for a crystalline undulator, where 
the number of channeling particles and the number of 
photons which can emerge from the crystal decrease with the growth of $\Lu$.
It is seen from (\ref{CUcharacteristics.6}) that 
if $\Lu \to \infty$ then the parameters $\kpd=\Lu/\Ld$
and $x\kpd=\Lu/\La$ also become infinitely large, and the 
effective number of periods goes to zero 
leading to $\Delta \cN_{\om_n}, \cF_n\to 0$.
This is quite natural result, since in the limit $\Lu\gg \Ld$
all particles leave the channeling mode and, thus, do not undulate
in the most part of the crystal, 
whereas all emitted photons are absorbed inside the crystal if
$\Lu\gg \La$.
Another formal (and physically trivial) fact, which follows
from  (\ref{CUcharacteristics.5}) and (\ref{CUcharacteristics.6}), 
is that $N_{\rm eff}(x,\kpd)=0$ also for a zero-length undulator,
when $\Lu=0$.
Vanishing of a positively defined quantity $N_{\rm eff}(x,\kpd)$
at two extreme boundaries suggests that 
there exists the length $\Lbar(x)$ 
for which the effective number of periods 
(taken for fixed values of $\La$, $\Ld$ and $\lamu$)
attains the maximum.

To define the value of $\Lbar(x)$ or, what is equivalent,
the quantity $\barkpd(x)=\Lbar(x)/\Ld$, one 
carries out the derivative of $f(x,\kpd)$ with respect
to $\kpd$ and equalizes it to zero.
The analysis of the resulting equation 
shows that for each value of $x=\Ld/\La\geq 0$ there is only
one root $\barkpd$.
Hence, the equation defines, in an inexplicit form, a single-valued
function $\barkpd(x)=\Lbar(x)/\Ld$
which ensures the maximum of $N_{\rm eff}(x,\kpd)$
for given $\La$, $\Ld$ and $\lamu$.

It is important to note that the crystal length enters 
Eqs.~(\ref{CUcharacteristics.4a})-(\ref{CUcharacteristics.4b}) only
via the ratio $\kpd$.
All other quantities, met in these formulae as well as in  
(\ref{CUcharacteristics.5}) and (\ref{CUcharacteristics.6}),
are independent on the length of the crystal.
Therefore, the quantity $\Lbar(x)$  
ensures the highest values  of $\Delta \cN_{\om_n}$ and $\cF_n$
for the radiation formed in the crystalline undulator.
In this sense  $\Lbar(x)$ can be called the {\it optimal length} 
of the undulator which corresponds to a given value of $x$.
\begin{figure}
\begin{center}
\includegraphics[width=9cm,height=9cm]{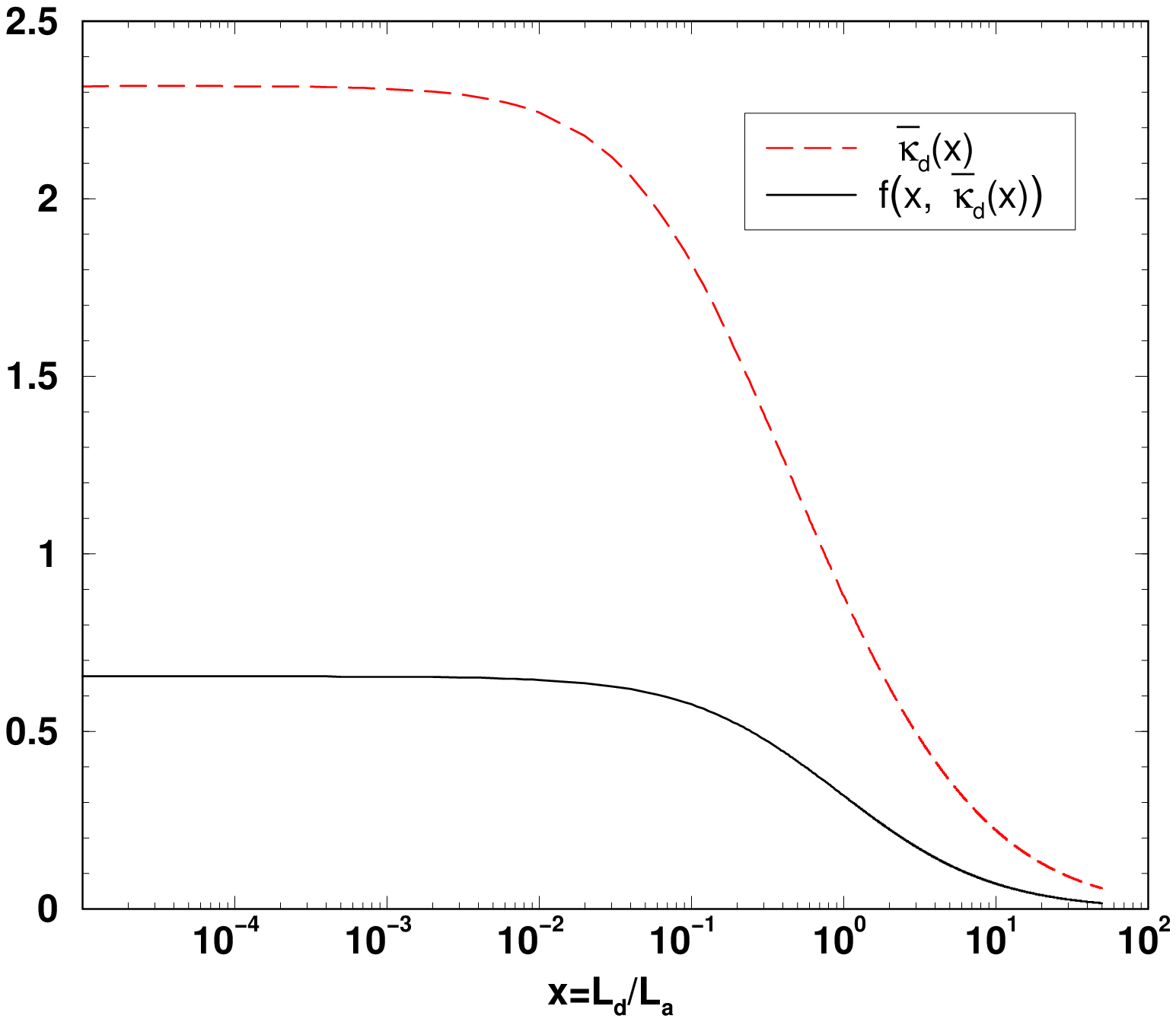}
\end{center}
\vspace{-0.5cm}
\caption{
Dependences $\barkpd(x)=\Lbar(x)/\Ld$ and 
$f_x(x,\barkpd(x)) = N_{\rm eff}(x,\barkpd(x))/\Nd$ 
on $x=\Ld/\La$.}
\label{OptimalUndNom.eps.fig}
\end{figure}

The dependences of $\barkpd(x)=\Lbar(x)/\Ld$ and of the ratio 
$f(x,\barkpd(x)) = N_{\rm eff}(x,\barkpd(x))/\Nd$
on $x$ are presented in Fig.~\ref{OptimalUndNom.eps.fig}.
For a given crystalline structure, the dechanneling length $\Ld$
is uniquely defined by the energy $\E$ and 
the parameters of bending $a$ and $\lamu$.
On the other hand, the attenuation length $\La$ is the function of 
$\om$. 
Therefore, fixing $\E$,  $a$, $\lamu$ and $\om$ one
calculates $x=\Ld/\La$ and, then, using the dashed curve 
in the figure finds the optimal length of the crystalline undulator 
$\Lbar(x)$ which accounts for the dechanneling effect and 
the photon attenuation.
Simultaneously, from the solid curve one finds the effective 
number of the undulator periods 
$N_{\rm eff}(x,\barkpd(x))$ which defines the number of emitted photons,
the flux and the brilliance of radiation.

\section{Numerical results}
\label{NumericalResults}

From (\ref{CUcharacteristics.4a}) follows,
that to find the number of photons $\Delta \cN_{\om_n}$ 
emitted in a crystalline one has to calculate two factors.
The factor $Q_n(p)$ (see (\ref{NumberPhotonsCollection.1})) 
depends on the harmonic number $n$ and on the 
undulator parameter $p$ which, in turn, is defined by the values of 
$\E$, $a$ and $\lamu$ through the relation $p=2\pi \gamma \, a/\lamu$.
The second factor, $N_{\rm eff}(x,\kpd)$ depends on $\Lu$, $\lamu$, 
$\Ld=\Ld(\E,C)$ and $\La=\La(\om)$.
It was explained in Sect. \ref{OptimalLength} that once the quantities
$\lamu$, $\Ld(\E,C)$ and $\La(\om)$ are known the length of the crystal 
can be fixed by the condition $\Lu=\Lbar$ which results in the maximum 
values of $N_{\rm eff}(x,\kpd)$ and $\cF_n$ 
with respect to $\Lu$.
The numerical data presented below in this section was obtained for 
the optimal length of undulator.

Therefore, to calculate $\Delta \cN_{\om_n}$ one fixes,
in addition to the crystallographic plane,
the values of $n$, $\E$, $a$ and $\lamu$ (the three latter
are subject to the conditions (\ref{Condition.1})--(\ref{Condition.4}))
which uniquely define the quantities $p$, $C$, $\om_n$, $\La(\om)$.
However, there is some uncertainty with respect to the magnitude of the
dechanneling length.
This uncertainty is not intrinsic to the case of a periodically bent crystal
but rather reflects the stochastic nature of the interaction of a 
channeling particle with crystal constituents. 
As mentioned above to calculate the dechanneling length one can apply 
the diffusion theory to describe the multiple 
scattering or carry out numerical simulations of the scattering 
process.
Alternatively, one can use model-dependent analytic expressions
for $\Ld(\E,C)$.
In the present paper we utilize the approach, presented in Ref. 
\citenum{KSG1999},
and  approximate $\Ld(\E,C)$ with
\begin{eqnarray} 
\Ld(\E,C) = (1-C)^2\, \Ld(\E,0),
\qquad
\Ld(\E,0)
=
{256 \over 9\pi^2}\,
{\atf\, d \over m c^2 \,r_{0}}\,
{\E \over \coul}
\, .
\label{DechannnelingAttenuation.3}
\end{eqnarray} 
Here 
where $r_{0}=2.8\times 10^{-13}$ cm is the electron classical radius,
$mc^2=0.511$ MeV is the electron rest energy,
$\atf$ is the Thomas-Fermi radius of the crystal atom.
The parameter $C$ is defined by Eq. (\ref{Condition.1}).
The quantity $\Ld(\E,0)$ stands for the dechanneling length of 
a positron in a straight crystal\cite{KSG01a,BiryukovChesnokovKotovBook}.
The quantity $\coul= \ln{\sqrt{2\gamma}mc^2/I} -23/24$, with  
$I$ denoting the (average) ionization potential of the crystal atom,
is the Coulomb logarithm characterizing the ionization losses of an
ultra-relativistic particle in amorphous media. 
For a quick estimation of $\Ld(\E,0)$ (in cm) one can re-write
the right-hand side of the second equation from
(\ref{DechannnelingAttenuation.3})
 as $2\atf\, d\,\E / \coul $, with
$\atf$ and $d$ measured in \AA\, and $\E$ in GeV.
The values of $\atf$ and $d$ , are presented in
Table~\ref{Crystals.Table}.  
\begin{table}[h]
\caption{
 Parameters $d$, $\atf$ and $\dUmax$ 
 for different crystals and channels.}
\label{Crystals.Table}
\begin{center}       
\begin{tabular}{|l|l|l|l|l|} 
\hline
                  & C (111) & Si (111) & Ge (111) & W (110) \\
\hline
$d$      (\AA)    & 1.54    & 2.35     &  2.45    & 2.24    \\
$\atf$   (\AA)    & 0.258   & 0.194    &  0.148   & 0.112   \\
$\dUmax$ (GeV/cm) & 9.23    & 8.58     &  17.5    & 57.4    \\
\hline
\end{tabular}
\end{center}
\end{table} 

To calculate the brilliance of a crystalline undulator
(which one obtains by using (\ref{CUcharacteristics.4b}) in 
(\ref{B&B.7}))
it is necessary to specify the parameters of a positron 
bunch, which are the current $I$, the beam sizes $\sigma_{x,y}$ and
angular divergencies $\phi_{x,y}$.
We used the parameters of the positron beams
from several modern high-energy $e^{-}e^{+}$ colliders. 
These parameters are summarized in 
Table~\ref{ParticleDataGroup2004.Table1}.
The data on  $\E$, $\sigma_{x,y}$, $l$, $\cN$ and  $I$
(which is an average beam current)
are taken from Ref.~\protect\citenum{ParticleDataGroup2004}.
The beam divergencies $\phi_{x,y}$ were calculated using
the data on the transverse emittance (not presented in the table) and
the beam size $\sigma_{x,y}$.
The peak current $I$, which is defined as the electric current 
of a single bunch,
 was calculated as  $I\, \mbox{(A)} \approx 48 \cN/l$ with $l$ in cm.
\begin{table}[h]
\caption{
Positron energy $\E$, bunch length $l$,
number of particles per bunch $\cN$,
beam sizes $\sigma_{x,y}$, 
beam divergencies $\phi_{x,y}$,
and a positron peak current $I$
for several modern high-energy $e^{-}e^{+}$ colliders 
\protect\cite{ParticleDataGroup2004}.}
\label{ParticleDataGroup2004.Table1}
\begin{center}       
\begin{tabular}{|l|l|l|l|l|l|l|} 
\hline
        &DA$\Phi$NE&VEPP-2000&BEPC-II &PEP-II&KEKB  &CERS-C \\
        &(Frascati)&(Russia)&(China)&(SLAC)&(KEK) &(Cornell)\\
\hline
$\E$ (GeV)  &0.700     &  1.0 &1.9-2.1&2.5-4 &3.5   & 6       \\
\hline
$l$ (cm)    &1-2       &   4  & 1.3   & 1    &0.65  & 1.2     \\
\hline
$\cN$ (units $10^{10}$) &3-9   &  16  & 4.8   & 6.7  & 7.3  &1.15\\
\hline
$\sigma_x$ (mm) & 0.800    &0.125 & 0.380 &0.157 &0.110 &0.300    \\
$\sigma_y$ (mm) & 0.0048   &0.125 & 0.0057&0.0047&0.0024&0.0057   \\
\hline
$\phi_x$ (mrad) & 0.375    & 2    & 0.379 &0.153 &0.164 &0.500    \\
$\phi_y$ (mrad) & 0.208    & 2    & 0.544 &0.319 &0.417 &0.439    \\
\hline
$I$ (A)         &144-216   & 192  & 177   & 322  &539   &46       \\
\hline
\end{tabular}
\end{center}
\end{table} 

\begin{figure}[h]
\begin{center}
\vspace*{-0.6cm}
\includegraphics[width=14.5cm,height=10.5cm]{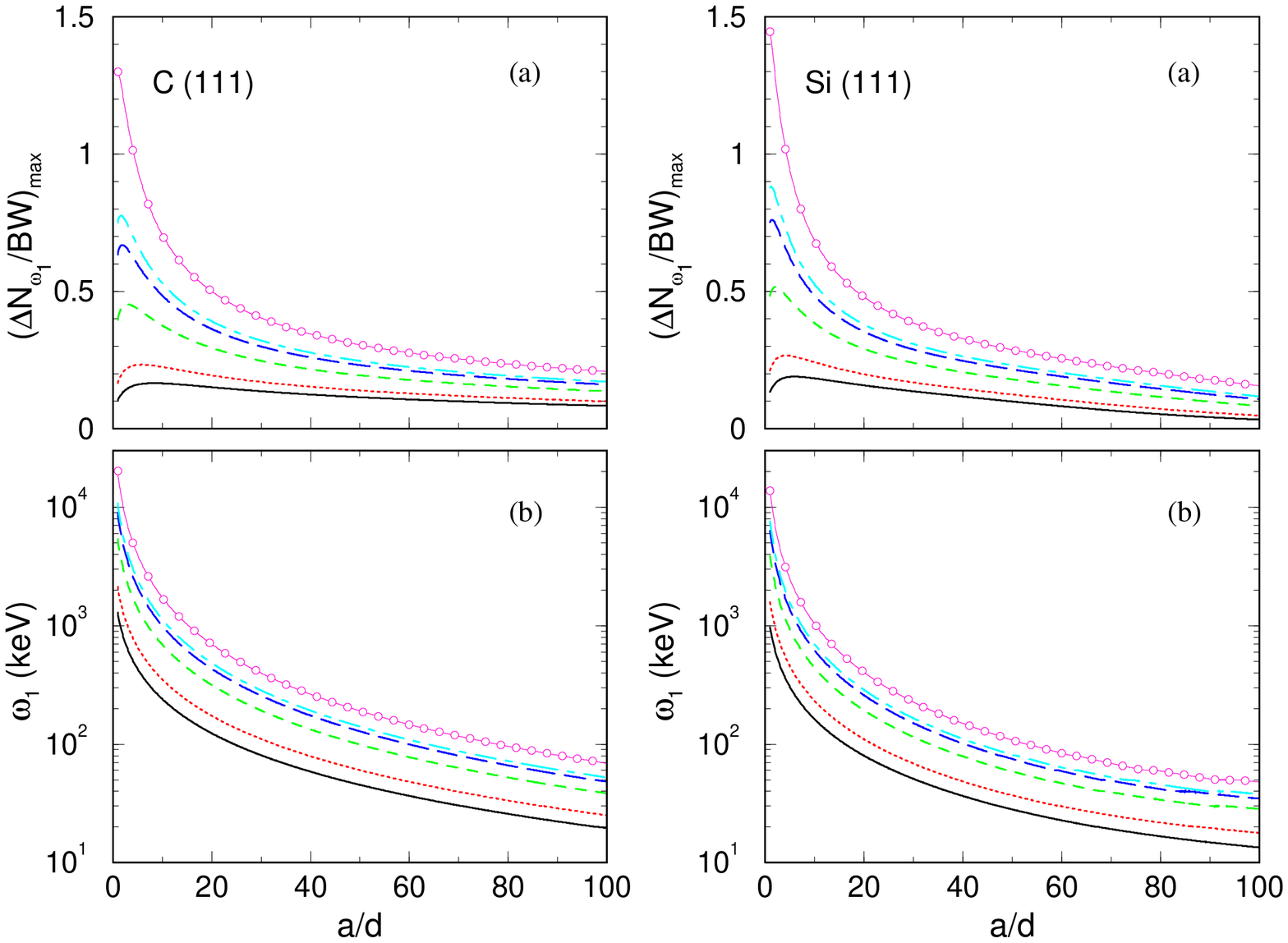}
\\
\includegraphics[width=14.5cm,height=10.5cm]{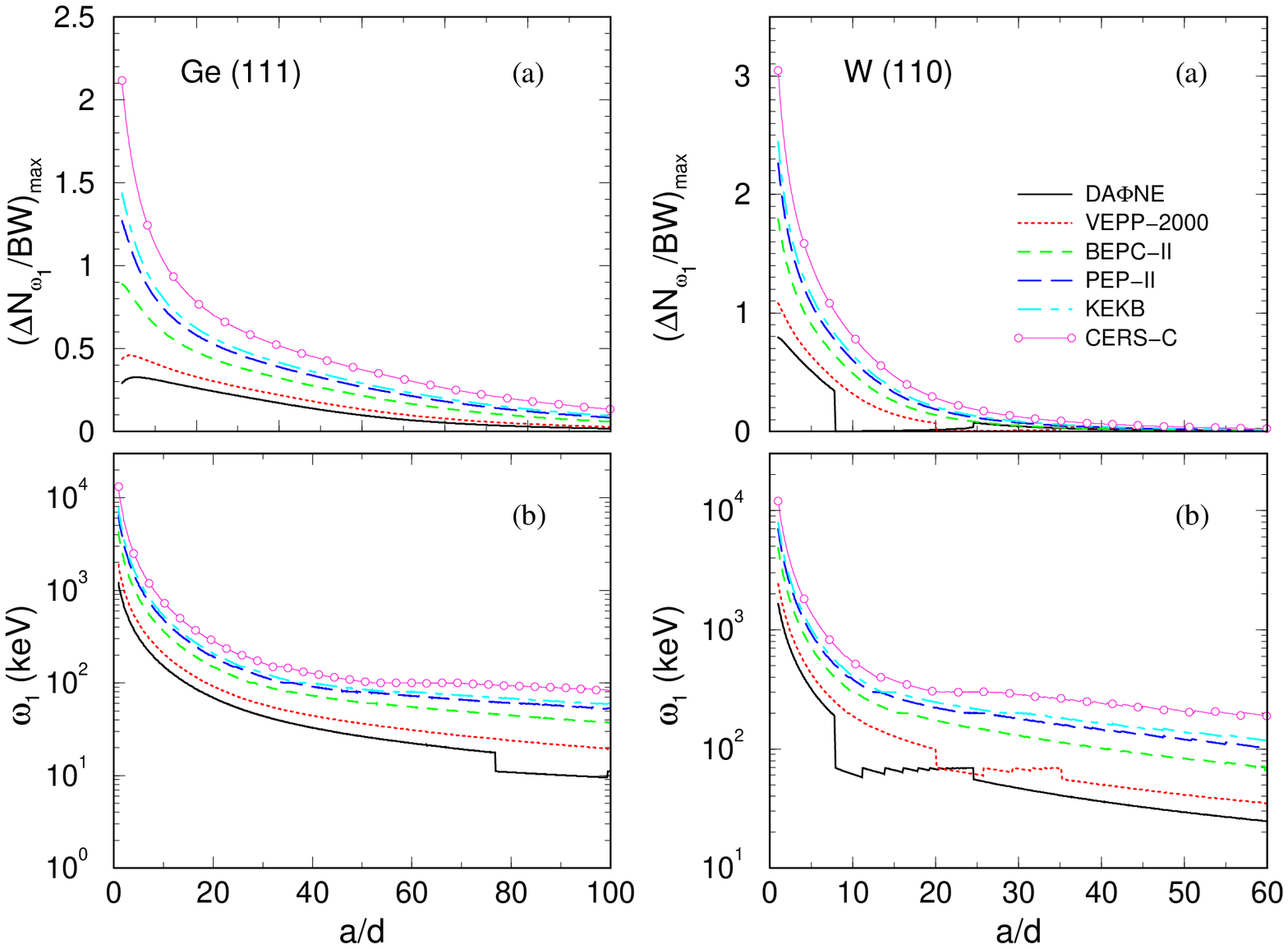}
\end{center}
\vspace{-1cm}
\caption{
Graphs (a):
the maximal number of photons of the first harmonic ($n=1$) 
per a bandwidth $\Delta \om_1/\om_1$ and per a positron
as a function of the ratio $a/d$ calculated for 
the positron energies in various colliders 
(see Table \ref{ParticleDataGroup2004.Table1})
as indicated.
Graphs (b): 
the corresponding values of the fundamental  harmonic energy
(see Eq. (\ref{CentralLine.1}) with $n=1$).
See also explanations in the text.
Each vertical pair of the graphs (a) and (b) correspond to the 
positron channeling in the particular periodically bent channel as indicated 
in the graphs (a).
The legend refers to all graphs in the figure.}
\label{S_Si_Ge_W.eps.fig}
\end{figure}

\begin{figure}
\begin{center}
\vspace*{-0.5cm}
\includegraphics[width=15cm,height=10cm]{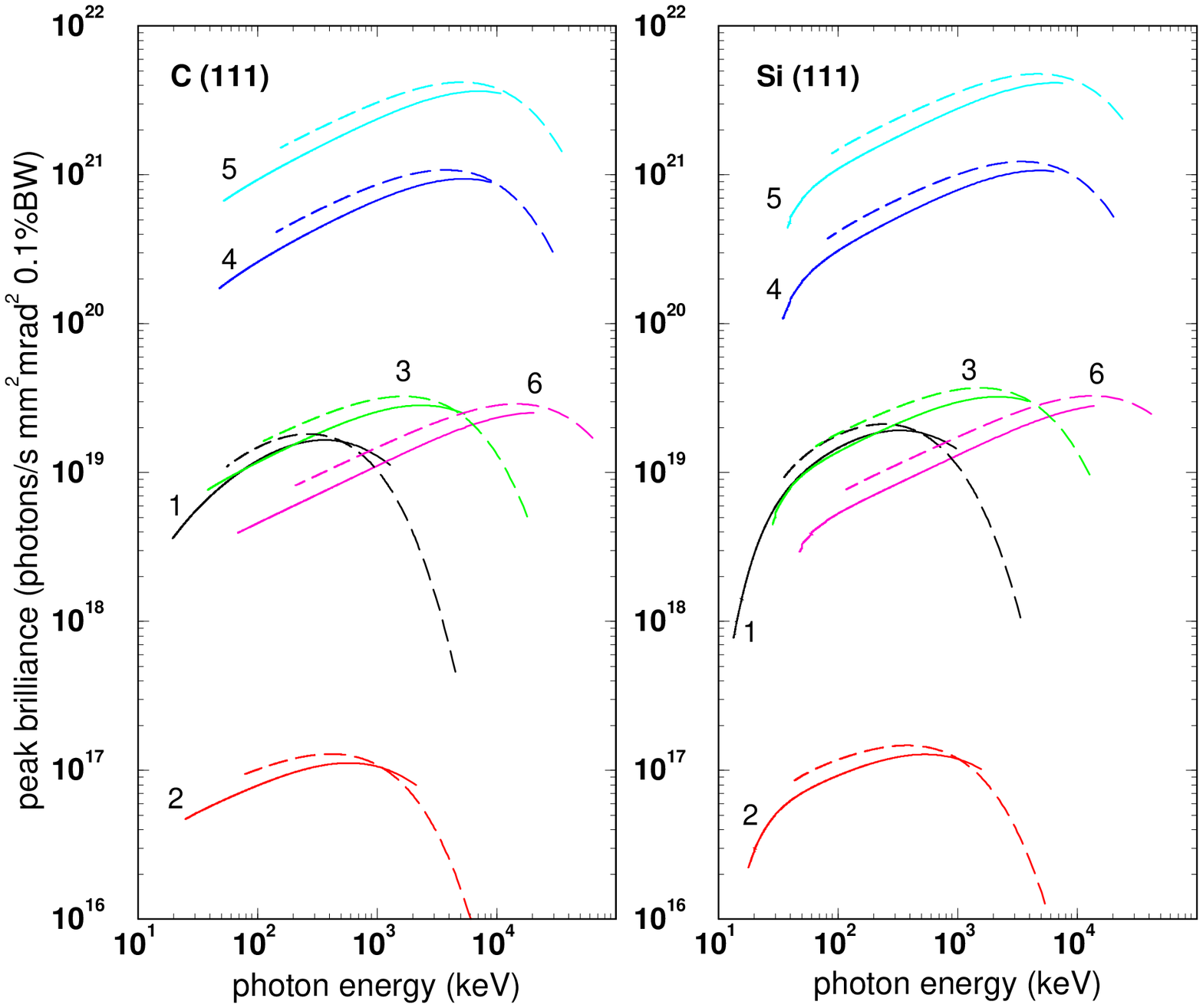}
\\
\includegraphics[width=15cm,height=10cm]{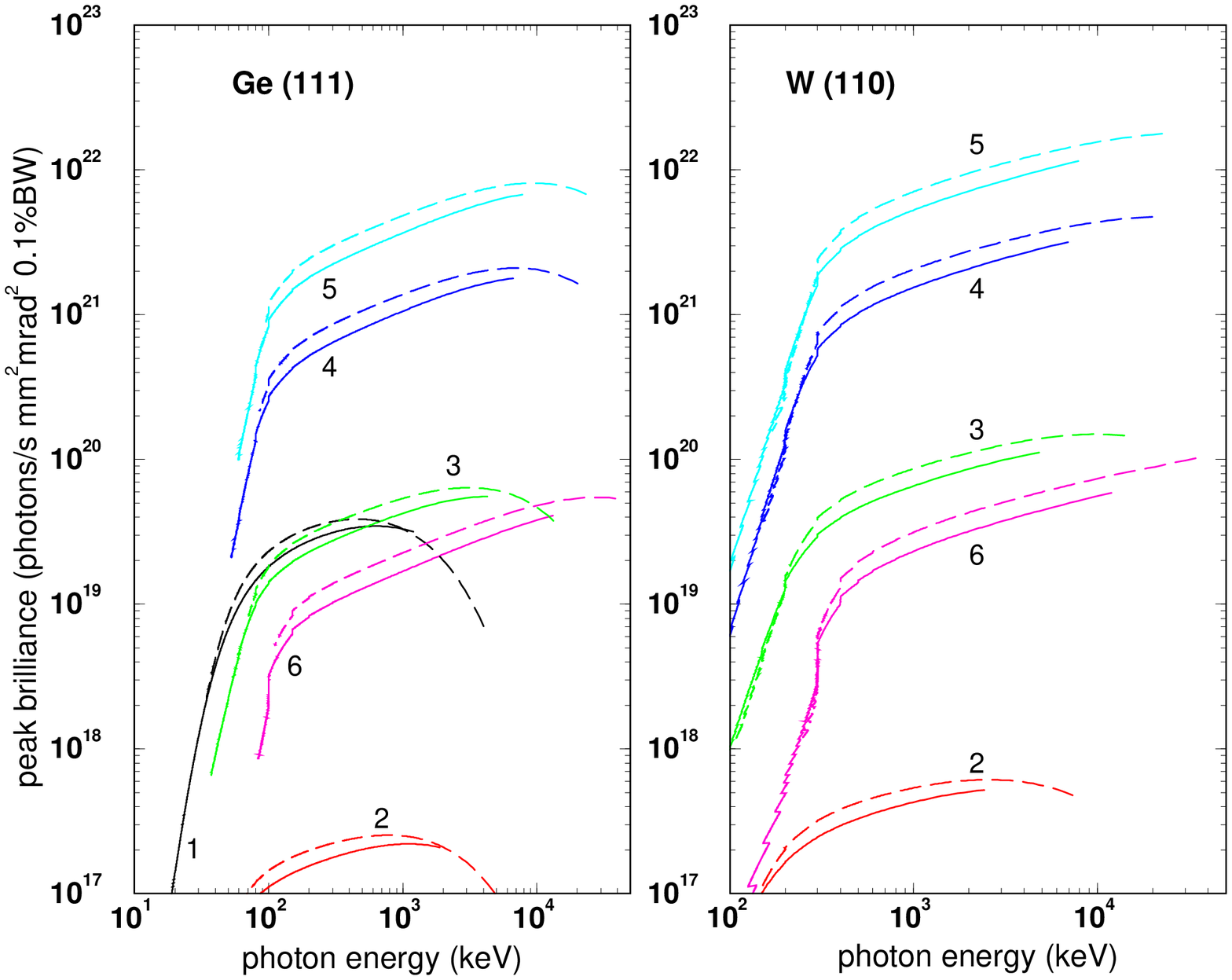}
\end{center}
\vspace{-0.5cm}
\caption{
Peak brilliance of the undulator radiation in the forward
direction calculated for
four channels as indicated in each graph.
The solid curves correspond to the radiation in the fundamental
harmonic $n=1$, the dashed curves refer to $n=3$.
In each graph the enumerated sets of the solid and the dashed curves 
correspond to the parameters of the positron beams in different colliders 
(see Table \protect\ref{ParticleDataGroup2004.Table1}).
1: DA$\Phi$NE, 2: VEPP-2000, 3: BEPC-II, 4: PEP-II, 5: KEKB,
6: CERS-C.}
\label{bril_C_Si_Ge_W.eps.fig}
\end{figure}

The results of our calculations are presented in 
Figs.~\ref{S_Si_Ge_W.eps.fig} and \ref{bril_C_Si_Ge_W.eps.fig}.
The choice of the crystals was motivated by the fact that C,
Si, Ge and W crystals are frequently used in 
channeling experiments 
(see, e.g., Ref.~\citenum{BiryukovChesnokovKotovBook}).
An additional reason is that for a given photon frequency 
the magnitude of $\La(\om)$ rapidly decreases with the growth
of atomic number of the constituent atoms.
Therefore, by comparing the results obtained for different crystals
one can investigate the influence of the photon attenuation
on the formation of the radiation in a crystalline undulator.

Graphs `(a)' in Fig.~\ref{S_Si_Ge_W.eps.fig} present the dependence 
of the {\it maximal} number of emitted photons of the first harmonic ($n=1$) 
per bandwidth $\Delta \om_1/\om_1$ and per a positron
versus the ratio $a/d$.
The curves were calculated for the positron energies 
indicated in Table \ref{ParticleDataGroup2004.Table1} 
(for BEPC-II and PEP-II colliders we used the values $\E=2$ GeV and
$\E=3$ GeV, respectively).
For each crystal and for each $\E$ value the dependences 
$\left(\Delta\cN_{\om_1} /\mbox{BW}\right)_{\max}$ 
were obtained as follows.
There are two independent variables, $\lamu$ and $a$, which, 
(for fixed crystal, energy and harmonic number $n$) define
all other quantities on the right-hand side of 
(\ref{CUcharacteristics.4a}).
For practical purposes it is more convenient to chose the 
ratio $a/d > 1$ and
the parameter $C<1$ (see Eq.~(\ref{Condition.1})) as 
the independent variables.
Then, for each pair $(a/d, C)$ one finds $\lamu$, $p=2\pi\,\gamma a/\lamu$, 
$Q_1(p)$,
the dechanneling length $\Ld(\E,C)$ and the number of periods 
$\Nd = \Ld(\E,C)/\lamu$,
the fundamental harmonic frequency $\om_1$ (see  Eq.~(\ref{CentralLine.1}))
and the attenuation length $\La(\om_1)$, and
the value of $N_{\rm eff}(x,\barkpd(x))$ which corresponds to the 
optimal undulator length calculated for  $x=\Ld(\E,C)/\La(\om_1)$
(see  (\ref{CUcharacteristics.5})-(\ref{CUcharacteristics.6}) and
Fig.~\ref{OptimalUndNom.eps.fig} and Sect.~\ref{OptimalLength}).
As a result, one finds the magnitude of $\Delta\cN_{\om_1} /\mbox{BW}$.
Finally, scanning through all $(a/d, C)$ values one determines the
highest possible value of the number of photons per BW,
$\left(\Delta\cN_{\om_1} /\mbox{BW}\right)_{\max}$, as a function of
$a/d$. 
Having done this one also finds the dependence $\om_1=\om_1(a/d)$
(graphs `(b)' in Fig.~\ref{S_Si_Ge_W.eps.fig})
as well all other characteristics of the undulator as functions
of $a/d$.

Let us briefly discuss the behaviour of obtained dependences.
Firstly, as it is seen from the graphs (a),
for a fixed amplitude $a$ the quantity
$\left(\Delta\cN_{\om_1} /\mbox{BW}\right)_{\max}$
is an increasing function of a positron energy $\E$.
This feature becomes clear if one analyzes 
the $\E$ dependence of the product $Q_1(p)\,\Nd\,f(x,\barkpd(x))$
which defines the number of emitted photons
(see Eqs.~(\ref{CUcharacteristics.4a}) and (\ref{CUcharacteristics.5})).
All three factors are increasing functions of energy
(although it is not too obvious for $f(x,\barkpd(x))$).

Another feature of the curves 
$\left(\Delta\cN_{\om_1} /\mbox{BW}\right)_{\max}$
is that they are decreasing function of $a/d$ in the region
$a/d>1$.
To a great extent this is a consequence of the photon attenuation
in the crystal.
Indeed, as the ratio $a/d$ increases the undulator period $\lamu$ increases
too, in order to maintain the inequality $C\ll 1$
(see Eq.~(\ref{Condition.1})).
Larger values of  $\lamu$ results in lowering of the 
emitted photon energy (see  Eq.~(\ref{CentralLine.1}) and the graphs
(b) in Fig.~\ref{S_Si_Ge_W.eps.fig}) and, consequently,
to the decrease of the attenuation length, $\La(\om)$.
This, in turn, leads to the increase  of the ratio $x=\Ld/\La$ which
defines the magnitude of $f(x,\barkpd(x))$.
This factor, as it is seen from  Fig.~\ref{OptimalUndNom.eps.fig},
rapidly falls off for $x>0.1$, and this feature manifests itself
in the dependence of $\left(\Delta\cN_{\om_1} /\mbox{BW}\right)_{\max}$
on $a/d$.
In the case of crystals consisting of heavy atoms the dependence 
acquires additional features, which are due to the fact that the 
ionization potentials, $I_0$, 
of the inner atomic subshells of such atoms lie
within the energy range $1\dots 100$ keV. 
The photons with the energy just above the threshold are absorbed
much more efficiently than those with the lower energies.
As a result, the dependence of $\La(\om)$ in the vicinity of the threshold
becomes a saw-like. 
For $\om < I_0$ the attenuation length noticeably (up to the order of
magnitude) exceeds $\La(\om)$ for $\om \geq I_0$.
This effect results in the irregularities of the dependence
$\left(\Delta\cN_{\om_1} /\mbox{BW}\right)_{\max}$ on $a/d$, which
in Figs.~\ref{OptimalUndNom.eps.fig} are mostly pronounced for diamond
and tungsten crystals.

In the opposite limit, when $a/d\ll 1$ the number of the emitted
photons goes to zero.
This tendency, which is seen explicitly for all the curves 
(but the CERS-C one) in the case of C and Si crystals, is also clear and
is due to the fact that the case $a=0$ corresponds to the linear
crystal, i.e. the absence of the crystalline undulator.

In Fig.~\ref{bril_C_Si_Ge_W.eps.fig} we present the peak brilliance
of the crystalline undulators based on different crystals (as indicated)
and calculated using the parameters of the positron beams from 
Table \protect\ref{ParticleDataGroup2004.Table1}.
The data refer to the emission in the first and the third 
harmonics in the forward direction.
It is seen that in contrast to the number of the emitted 
photons which is the same, by the order of magnitude, for all 
colliders, the magnitudes of the peak brilliance for different 
beams differ by orders of magnitude.
To the largest extent this is due to the quality of the beam,
which includes, apart from the beam current $I$, its 
size and angular divergency, see. Eqs.~(\ref{B&B.4}) and
(\ref{B&B.7}).
For all crystals and over the whole range of photon energies the
product $\epsilon_x\epsilon_y$ of the photon source emittances 
 is the smallest for the KEKB collider (labeled as `5' in the graphs
in Fig.~\ref{bril_C_Si_Ge_W.eps.fig}).
As a result, this beam, which does not lead to the highest
values of $\left(\Delta\cN_{\om_1} /\mbox{BW}\right)_{\max}$,
ensures the largest peak brilliance of the crystalline undulator
radiation.
The peak brilliance for the KEKB positron beam is on the level of 
$(4\dots 20)\times 10^{22}\,
\Bigl(\mbox{photons}/\mbox{s}/\mbox{mrad}^2/\mbox{mm}^2/0.1\%\mbox{BW}\Bigr)$ 
for the photon energies within $1\dots 10$ MeV range.
These values can be compared with the peak brilliance of the
light sources of the third generation \cite{EuroPhys}.
The peak brilliance on the level $10^{21}\dots10^{23}$ in the 100 keV 
range of photon energies by means of 
the undulators based on the action of magnetic field 
is planned to be achieved within several projects 
\cite{WebPage,Petra3,Tesla}.
The data from Fig.~\ref{bril_C_Si_Ge_W.eps.fig}) demonstrate that
it is feasible to produce the radiation of the same level of
brilliance but for much higher energies by means of crystalline undulators.

\section{Conclusion}
\label{Conclusion}

Theoretical investigations show that it is entirely 
realistic to use a crystalline 
undulator for generating spontaneous radiation in a wide range
of photon energies.
The parameters of such an undulator, being subject
to the restrictions mentioned in Sect.~\ref{Introduction}, can be 
easily tuned by varying the parameters of the bending, the positron energy 
and by choosing different channels.
The large range of energies available in modern colliders 
together with the wide range preparation of periodically bent crystalline
structures allow one 
to generate the crystalline undulator radiation 
with energies from hundreds of keV up to tens of MeV region. 
The brilliance of the undulator radiation within this energy range
is comparable to that of conventional light sources of the third
generation but for much lower photon energies.

The experimental efforts are needed for the verification of  numerous 
theoretical predictions. 
Such efforts will certainly make this field of 
endeavor  even more fascinating than as it is already and 
will possibly lead to the practical development of
a new type of tunable and monochromatic radiation  sources.

\acknowledgments    

AVK acknowledges the support from the Alexander von Humboldt Foundation.


\end{document}